\documentclass[12pt,preprint]{aastex}
\usepackage{amssymb, amsmath}
\usepackage{color}
\usepackage{morefloats}
\usepackage{hyperref}

\shorttitle{A Runaway Yellow Supergiant Star in the Small Magellanic Cloud}

\shortauthors{Neugent et al.}

\begin{document}

\title{A Runaway Yellow Supergiant Star in the \\Small Magellanic Cloud\altaffilmark{*}}

\author{Kathryn F.\ Neugent\altaffilmark{1,2}, Philip Massey\altaffilmark{1,3}, Nidia Morrell\altaffilmark{4}, Brian Skiff\altaffilmark{1}, and Cyril Georgy\altaffilmark{5}}

\altaffiltext{*}{This paper includes data gathered with the 6.5-m Magellan Telescopes located at Las Campanas Observatory, Chile.}
\altaffiltext{1}{Lowell Observatory, 1400 W Mars Hill Road, Flagstaff, AZ 86001; kneugent@lowell.edu; phil.massey@lowell.edu; bas@lowell.edu}
\altaffiltext{2}{Department of Astronomy, University of Washington, Seattle, WA, 98195}
\altaffiltext{3}{Department of Physics and Astronomy, Northern Arizona University, Flagstaff, AZ, 86011-6010}
\altaffiltext{4}{Las Campanas Observatory, Carnegie Observatories, Casilla 601, La Serena, Chile; nmorrell@lco.cl}
\altaffiltext{5}{Geneva University, Geneva Observatory, CH-1290 Versoix, Switzerland; cyril.georgy@unige.ch}

\begin{abstract}
We recently discovered a yellow supergiant (YSG) in the Small Magellanic Cloud (SMC) with a heliocentric radial velocity of $\sim300$ km s$^{-1}$ which is much larger than expected for a star in its location in the SMC. This is the first runaway YSG ever discovered and only the second evolved runaway star discovered in a different galaxy than the Milky Way. We classify the star as G5-8~I, and use de-reddened broad-band colors with model atmospheres to determine an effective temperature of $4700\pm 250$K, consistent with what is expected from its spectral type. The star's luminosity is then $\log L/L_\odot \sim 4.2 \pm 0.1$, consistent with it being a $\sim$ 30Myr 9$M_\odot$ star according to the Geneva evolution models. The star is currently located in the outer portion of the SMC's body, but if the star's transverse peculiar velocity is similar to its peculiar radial velocity, in 10~Myr the star would have moved 1.6$^\circ$ across the disk of the SMC, and could easily have been born in one of the SMC's star-forming regions. Based on its large radial velocity, we suggest it originated in a binary system where the primary exploded as a supernovae thus flinging the runaway star out into space. Such stars may provide an important mechanism for the dispersal of heavier elements in galaxies given the large percentage of massive stars that are runaways. In the future we hope to look into additional evolved runaway stars that were discovered as part of our other past surveys.
\end{abstract}


\section{Discovery}
\begin{figure}
\epsscale{1}
\plotone{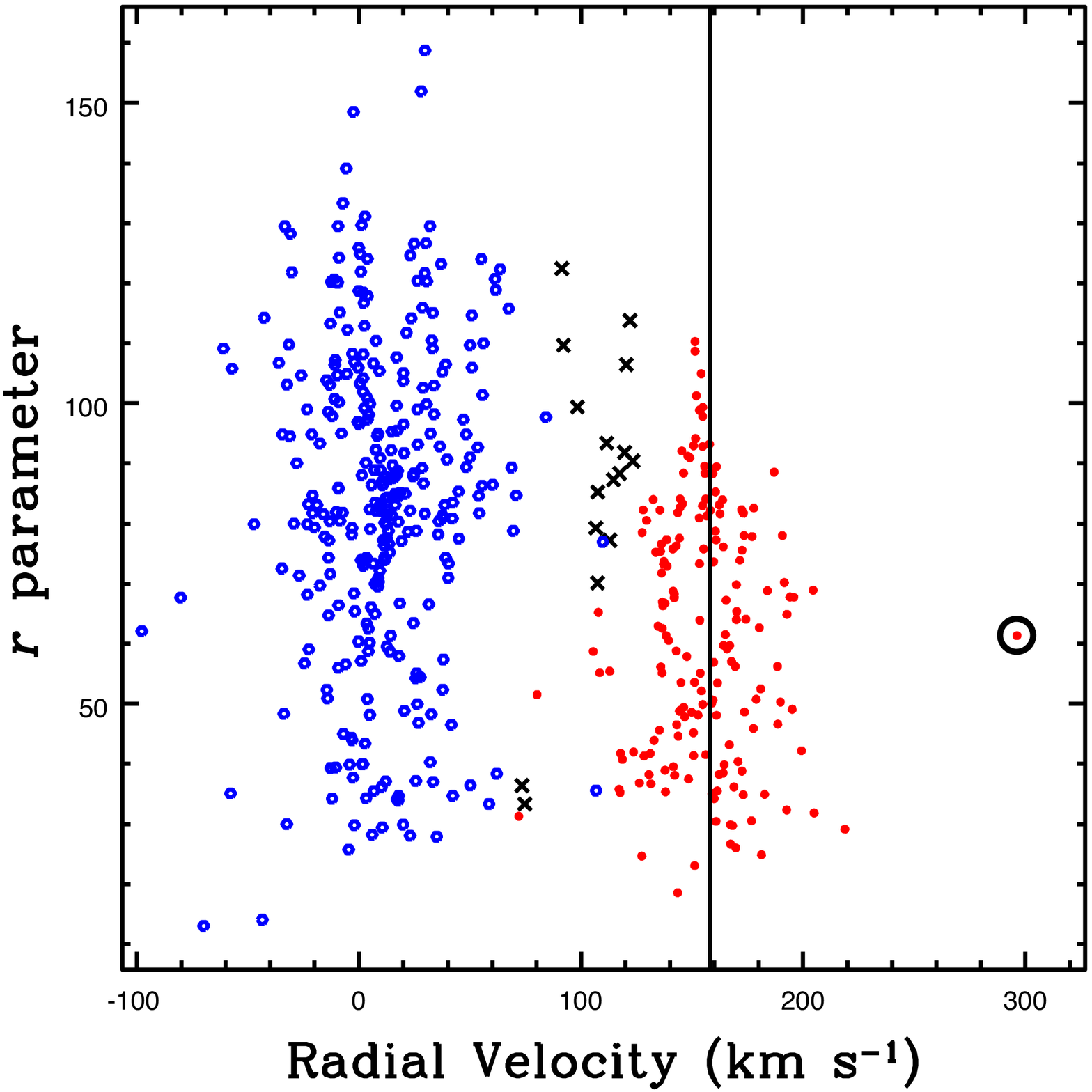}
\caption{\label{fig:VelR} Separation of Foreground and SMC Yellow Stars. Blue open hexagons represent foreground yellow dwarfs. Notice how they're clustered around 0 km s$^{-1}$. Red filled circles represent SMC yellow supergiants as they're clustered around the SMC's average heliocentric radial velocity of 158 km s$^{-1}$. Black $\times$s in the middle represent stars with uncertain classifications. The y axis is the Tonry and Davis (1979) $r$ parameter which is a proxy for the error on each measurement with high values indicating lower errors. This paper discusses the star with the abnormally large radial velocity located at $\sim300$ km s$^{-1}$. Data from Neugent et al. (2010) with a small revision as described in the text.}
\end{figure}

Neugent et al.\ (2010) conducted a radial velocity study of yellow stars seen in the direction of the Small Magellanic Cloud (SMC) in order to identify its yellow supergiant (YSG) population. As shown in Figure~\ref{fig:VelR}, the observed radial velocities are either clustered around 0 km s$^{-1}$ (as expected for foreground yellow dwarfs), or around the SMC's heliocentric radial velocity of 158 km s$^{-1}$ (Richter et al.\ 1987) as indicated by the black line and expected for SMC yellow supergiants. One star, J01020100-7122208, however, has a heliocentric radial velocity of around 300 km s$^{-1}$, 140 km s$^{-1}$ greater than expected. Neugent et al.\ (2010) don't explicitly comment on this star; at the time we believed it to be a likely short-period binary. However, we have now completed additional observations that rule out this explanation. Instead, this star is the first runaway YSG discovered and the second evolved runaway star discovered in another galaxy.

The concept of runaway massive stars has been around for sixty years. Blaauw (1956a, 1956b) discovered that some OB stars stars have much higher space velocities than other stars in their surroundings. Zwicky (1957) was the first to hypothesize that such stars should exist due to supernovae explosions causing the secondary in a binary system to be shot off into space. However, Gies \& Bolton (1986) showed that the close binary frequency of runaway OB stars was the same as for non-runaways, suggesting that the binary ejection mechanism was not the primary source of runaway stars. There have since been other theories as to how such runaway stars exist including the effects of different dynamical interactions (Leonard \& Duncan 1990) or interactions with massive black holes (Capuzzo-Dolcetta \& Fragione 2015; Fragione \& Capuzzo-Dolcetta 2016). Runaway stars aren't rare, with as many as 50\% of OB stars considered runaways with peculiar (difference between observed and expected) radial velocities larger than 40 km s$^{-1}$ (Gies \& Bolton 1986). 

Given that such a large fraction of OB stars are runaways, it is surprising that there are very few evolved massive stars known to be runaways.  Three Galactic red supergiants (RSGs), Betelgeuse, $\mu$ Cep and IRC-10414, are thought to be runaways based upon the presence of bow shocks (Noriega-Crespo et al.\ 1997, Cox et al.\ 2012, and Gvaramadze et al.\ 2013).  Evans \& Massey (2015) recently identified a runaway RSG in the Andromeda Galaxy directly from its peculiar velocity. The star we discuss here is the first known runaway YSG anywhere and is only the second known evolved massive star runaway of any kind found in another galaxy.  In Section 2 we'll discuss our observations and how we calculated the radial velocities. In Section 3, we'll give an overview of the physical properties of this runaway YSG. In Section 4 we'll discuss our results and conclusions.

\section{Observations, Reductions, and Radial Velocity Calculation}
As mentioned above, we originally believed that the star's abnormally large radial velocity was due to binary motion. Once we entertained the idea of it being a runaway star we obtained three additional spectra to investigate this possibility. 

Our first (discovery) spectrum was obtained using Hydra on the CTIO 4-m Blanco telescope on (UT) 2009 Oct 9.  The second spectrum was obtained on (UT) 2017 Aug 16 using the Echelle on the du Pont 2.1-m telescope on Las Campanas in Chile.  The spectrum was under-exposed in the blue, as our goal was simply to check on the radial velocity from the Ca II triplet in the far red.  Subsequently (2017 Dec 31) we obtained a high signal-to-noise spectrum with MagE on the Las Campanas Baade 6.5-m Magellan telescope for the purposes of spectral classification and radial velocity measurement, plus a shorter exposure the following night simply to confirm the radial velocity.  The wavelength ranges, spectral resolution ($\Delta \lambda/\lambda$), and exposure times are summarized in Table~\ref{tab:obs}. To help with spectral classification we also observed six bright ($V\sim 10-12$) yellow supergiants in the SMC and LMC, ranging in type from F0 to G8.

The full calibration and reduction details for the Hydra spectrum are given in Neugent et al.\ (2010).  For the MagE spectra, we obtained bias and flat-field exposures during the day, and each program exposure was followed immediately by a comparison arc exposure. We used Jack Baldwin's {\it mtools} routines in IRAF\footnote{IRAF is distributed by the National Optical Astronomy Observatory, which is operated by the Association of Universities for Research in Astronomy (AURA) under a cooperative agreement with the National Science Foundation.} for the spectral extractions; these routines are available from the Las Campanas website. The standard IRAF echelle reduction tasks were then used for wavelength calibration and flux calibration using spectrophotometric standards. For the duPont data, Milky flats were obtained with the afternoon sky and a diffusor in front of the slit, and the reductions are all performed with IRAF, without intervention of the 'mtools' package.

We measured the radial velocities using two methods: first, by fitting a Gaussian to each of the Ca\,{\sc ii} triplet lines ($\lambda 8498$, $\lambda 8542$, $\lambda 8662$), and secondly by cross-correlating the Hydra radial velocity standard stars from Neugent et al.\ (2010) against the new spectra in the region around the Ca\,{\sc ii} triplet.  The two methods agreed to within 1 km s$^{-1}$.  We give the average heliocentric radial velocities (HRV) of the two methods in Table~\ref{tab:obs}. All four values are consistent with each other, essentially ruling out the possibility that the large peculiar radial velocity is due to motion in a close binary system. A small difference is seen in the values between various telescopes, which we believe is due to minor offsets in the instrumental velocity zero-points.  To investigate this further, we measured the radial velocities of the six MagE spectral standards, five of which have published radial velocities from other sources. As is shown in Table~\ref{tab:stds}, we find that the differences ranged from -14.0 km s$^{-1}$ to +18.6 km s$^{-1}$, consistent with our assertion that our measurements do not indicate any significant radial velocity variations for our runaway. 

For consistency we also remeasured the velocity from our original Hydra spectrum. Much to our chagrin, we discovered that we had originally misapplied the heliocentric correction. This affects all of the individual radial velocities given in Table~1 of that paper by $\sim$11.0 km s$^{-1}$, with the previously published values too large. We take the opportunity to reissue the table here, as Table~\ref{tab:allobs}.  Note, however, that this does not affect anything else in that paper, as our assignment of membership to the SMC was based upon the distribution of velocities, and those were completely accurate in a relative sense, since the heliocentric corrections for all stars were the same. We also took the opportunity to update the table with revised spectral types that were subsequently published.

\section{Physical Properties}
In this section we use our spectral information and photometry to determine the star's effective temperature and luminosity, allowing us to place the star on the H-R diagram. We can then use evolution models to determine other physical properties, such as approximate mass and age.

\subsection{Spectral Classification}
\begin{figure}
\epsscale{1}
\plotone{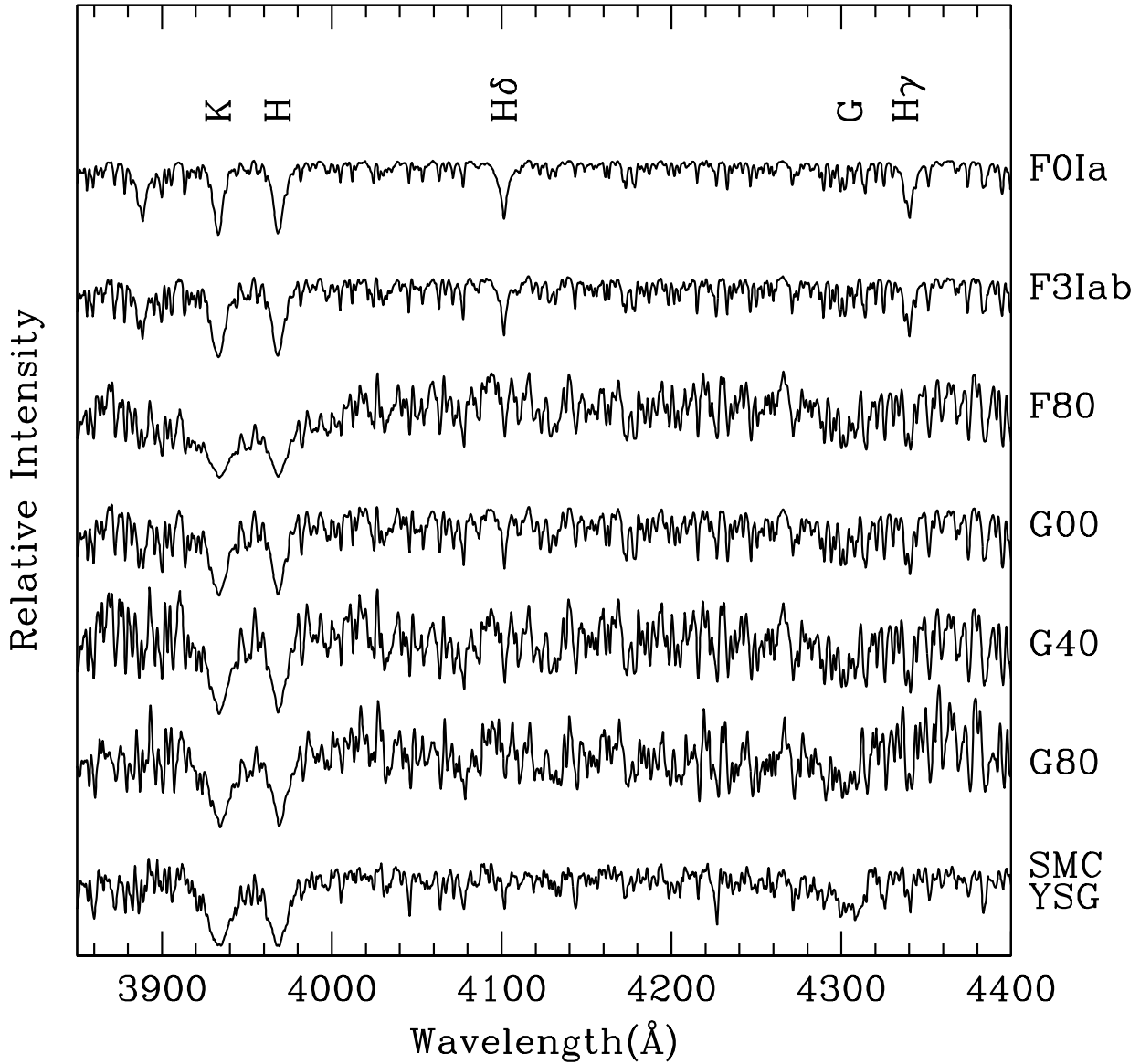}
\caption{\label{fig:type} Spectral classification stars and runaway star spectrum. We classified the runaway YSG by comparing it to other LMC and SMC YSGs with well-established types. The prominent G-band, abundance of metal lines and lack of hydrogen lines point to it being a G5-8~I. The spectrum of the YSG is shown on the bottom. The stars shown are the spectral standards Sk 105 (F0 Ia), Sk 55 (F3 Iab), HD 271182 (F8 0), HD 269953 (G0 0), HD 269723 (G4 0), HD 268757 (G8 0), along with our YSG runaway, J01020100-7122208. We are making these data available electronically as the ``Data behind the Figure."}
\end{figure}

In Figure~\ref{fig:type} we compare the spectrum of our star to six spectral standards, Sk 105 (F0 Ia), Sk 55 (F3 Iab), HD 271182 (F8 0), HD 269953 (G0 0), HD 269723 (G4 0), HD 268757 (G8 0).  The two early F stars are SMC members classified by Ardenberg \& Maurice (1977); the others are LMC members classified by Keenan \& McNeil (1989), and considered to be MK standards.  (The luminosity class ``0" is one notch above the Ia designation.)   We see immediately that our runaway is of G-type, based both upon the abundance of metal lines (despite the lower metallicity of the SMC compared to that of the LMC), and the weakness of the hydrogen lines.  The strength of the G-band argues that the star is closer to type G8 than to G0; we assign a type of G5-8~I. For G stars, the strengths of the Ca II H and K lines provide a crude luminosity indicator, developing wide wings at high luminosities (Kaler 2011), and it is clear from Figure~\ref{fig:type} that these lines in our runaway are comparable to those of the supergiant standards.

\subsection{Effective Temperature}
The uncertainties in the effective temperature scale for late-type supergiants are nicely discussed in the classic review by B\"{o}hm-Vitense (1981). A typical temperature for a G8~I star is 4570-90~K, while a G3~I star would be 4980~K (B\"{o}hm-Viense 1981, Cox 2000). Thus the expected temperature range for our star would be 4500-4800~K, with the caveat that the relationship between spectral type and effective temperature has only been established for solar metallicity stars. Since the metallicity of the SMC is roughly one-quarter solar, we expect that the effective temperature will be slightly cooler, probably by $\sim100$K, based on our previous modeling experience of RSGs (see e.g., Levesque et al.\ 2006).

We can refine our estimate of the effective temperature by instead using broad-band photometry and appealing to stellar atmosphere models of the appropriate metallicity.  The major uncertainty is what to assume for the reddening.   The observed colors are included in Table~\ref{tab:allobs}.   According to Cox (2000), the intrinsic $(B-V)_0$ of a G5~I star is 1.02, while that of a G8~I is 1.14. The observed $B-V$ color is 1.15, suggesting a color excess in the range $E(B-V)=0.01-0.14$.  However, we know that the foreground reddening towards the SMC corresponds to an $E(B-V)=0.04$, while the typical color excess of SMC OB stars is $E(B-V)=0.09$ (Massey et al.\ 2007).   Let us consider then three different values for the reddening,  $E(B-V)=0.04$ (minimum, only foreground), 0.09 (typical of OB stars), and 0.14. For the other colors, we adopt the relations between color excesses given by Schlegel et al.\ (1998), e.g., $E(U-B)=0.72E(B-V)$, $E(V-K)=2.95E(B-V)$, and $E(J-K)=0.54E(J-K)$.

In Table~\ref{tab:temps} we list the deredded colors, along with the corresponding effective temperatures derived from the updated ATLAS9 models (Castelli \& Kurucz 2003) for a metallicity of [-0.5], corresponding to 0.3$\times$ solar, appropriate for the SMC\footnote{\url{http://www.user.oats.inaf.it/castelli/colors/BCP/BCP_m05k2odfnew.dat}}.  The range of reddening we consider changes the derived temperature by only 100~K.  There is a larger difference (500~K) depending upon which color index we use\footnote{To compute the temperature from the models we applied simple linear interpolation between the colors and temperatures.  Using the models required us to adopt a value for the surface gravity, $g$.  We had to perform this process iteratively, as the surface gravity depends upon the adopted luminosity to compute the stellar radius, and luminosity depends upon the temperature for the bolometric correction.  We assumed a mass of $9M_\odot$ based upon the stellar evolution tracks described in the next section. This process converged quickly, with a $\log g$ of 0.8 [cgs].}. We adopt an average temperature 4700 +/- 250, in good agreement with the temperature we obtained via the spectral type.  Note that the two methods are essentially independent of each other.

We adopt an absolute visual magnitude of the star $M_V=-5.4\pm0.2$, based upon a distance of the SMC of 59~kpc (van den Bergh 2000).  The uncertainty represents the difference in the correction for extinction based upon the amount of reddening assumed, with $A_V=3.1E(B-V)$.  The bolometric correction is then $-0.4\pm0.2$ based on the ATLAS9 models, with the error corresponding to the uncertainty in temperature.  Thus the star's luminosity is $\log L/L_\odot=4.2\pm0.1$.  We summarize the physical properties in Table~\ref{tab:physprop}.

\subsection{H-R Diagram}
\begin{figure}
\epsscale{1}
\plotone{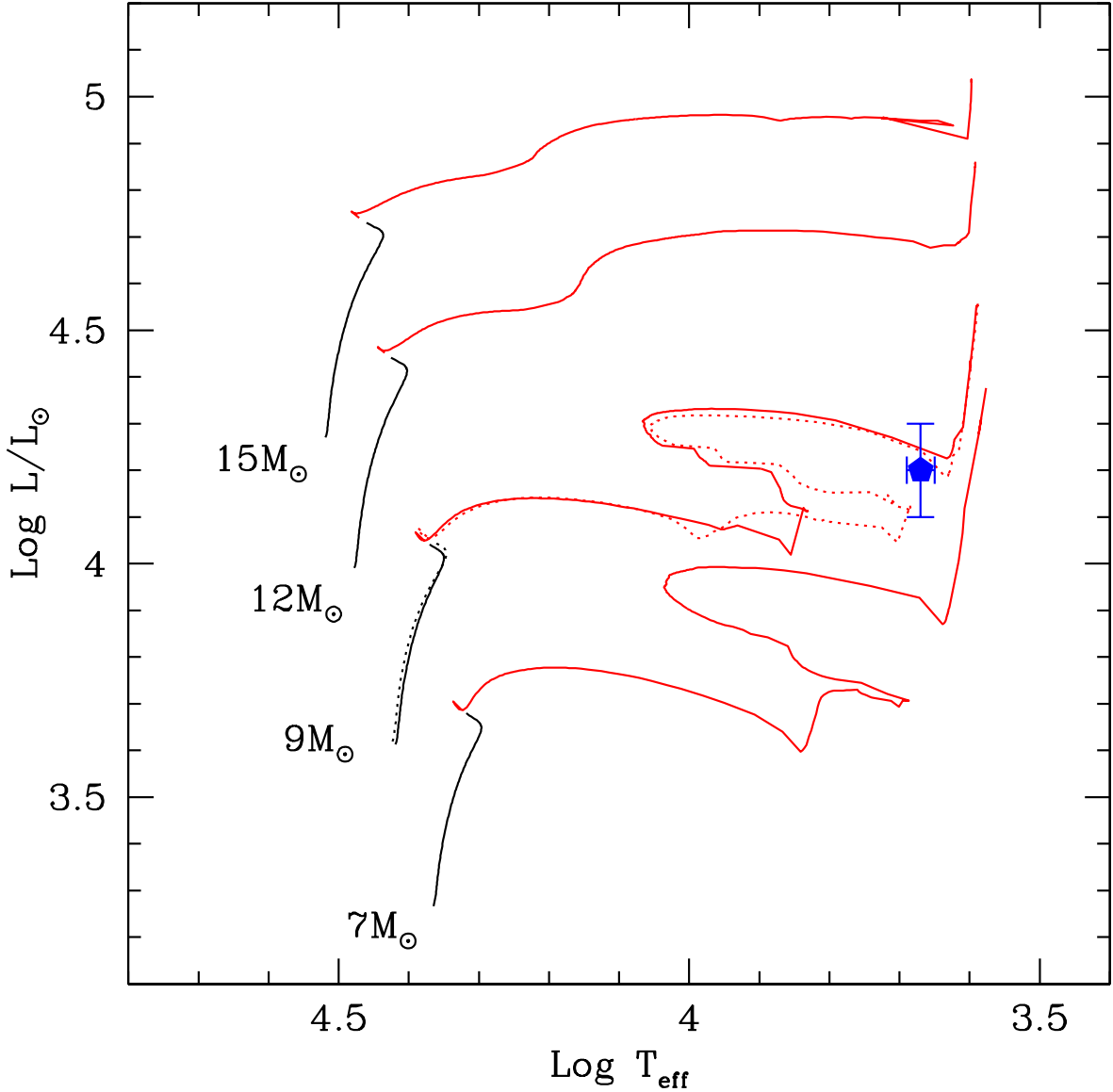}
\caption{\label{fig:HRD} H-R diagram along with Geneva evolution tracks. The luminosity and temperature are shown with errors. The Geneva z=0.006 evolution tracks are plotted in black during the main-sequence phase and then red during the evolved stages. These tracks were interpolated to a metallicity of z=0.004 (SMC-like) using the interactive web tool at the Geneva web site (\url{https://obswww.unige.ch/Recherche/evoldb/index/Interpolation/}) starting with the B-type stars grid published by Georgy et al.\ (2013) and adopting an initial equatorial rotation velocity of 40\% of the breakup speed ($\Omega / \Omega_{\rm crit}=0.568$). The dotted 9$M_\odot$ track is for a value of $\Omega / \Omega_{\rm crit}=0.379$.  Regardless of the assumption, the tracks indicate a mass of 9$M_\odot$ and an age of about 30~Myr for the YSG runaway.}
\end{figure}

In Figure~\ref{fig:HRD} we show the star's location in the H-R diagram and include the latest version of the Geneva evolution tracks. These have been interpolated to $z=0.004$ (appropriate for the SMC) using the grid of models from Georgy et al. (2013). We see that the luminosity and effective temperature of our runaway corresponds to a star with an initial mass of about 9$M_\odot$. The tracks were computed using the assumption of an initial equatorial rotational velocity that is 40\% of the critical break-up speed, which is probably the most representative (see discussion in Ekstr\"{o}m et al.\ 2012 and Georgy et al. 2013). However, we show by the dotted 9 $M_\odot$ track the effect of a 1.5$\times$ lower rotational velocity. The age associated with this stage is about 30~Myr, regardless of the initial assumptions. It must be noted, however, that the expulsion mechanism behind this star becoming a runaway could have altered its evolutionary background and thus the age and mass we are able to predict.

\section{Discussion and Conclusions}

We briefly considered the possibility that this was was a foreground object: a very metal poor star out in the Galactic halo, for instance, could have a very high space velocity. Once we obtained the blue/optical spectrum described here, though, we could rule out this interpretation, as such a metal-poor object would show almost no spectral features other than the Balmer lines, the G-band, and H and K, while our optical spectrum discussed above shows numerous metal lines.  The remote possibility that this was a foreground dwarf with an extraordinary radial velocity could also be dismissed from the wide wings of of the H and K lines, indicating high luminosity, as discussed in Section 3.

\begin{figure}
\epsscale{1}
\plotone{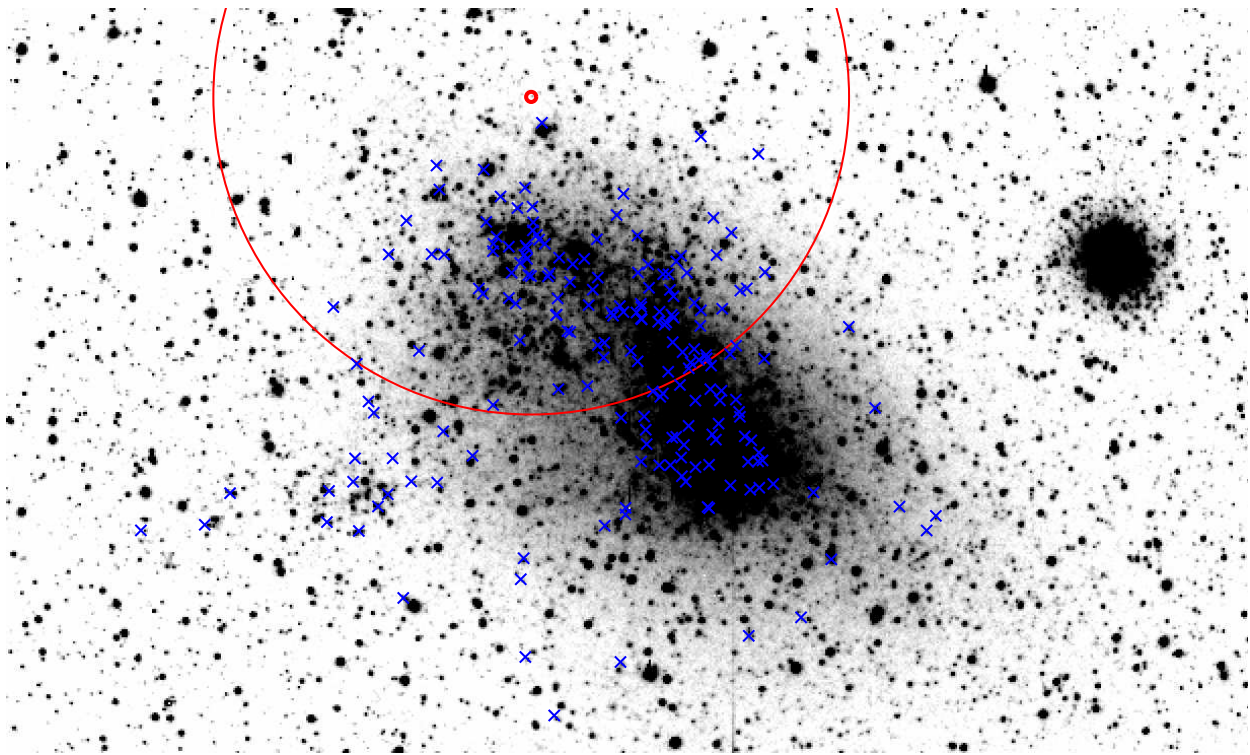}
\caption{\label{fig:location} Location of the runaway YSG within the SMC. The location is indicated by the small red circle while the location of the other SMC YSGs are shown as blue $\times$s. The large red circle denotes the 1.6$^{\circ}$ the star would travel over 10 million years if its transverse velocity is similar to its radial peculiar velocity.}
\end{figure}

The spatial location of the runway YSG is shown in Figure~\ref{fig:location}. It is not located within the central portion of the SMC and is instead out on the edge of the galaxy, far from any regions of current star formation. We posed the following question: If the star's tangential motion is the same as its peculiar radial velocity, how far would it move during its lifetime? We used 10~Myr as a ball-park estimate for the time involved, since we do not know how long it has been traveling (the star's age is 30 Myr); in that period of time it would have moved 1.6 degrees! This distance is shown in Figure~\ref{fig:location} as the large red circle. It clearly encompasses the central portion of the SMC suggesting that this star could have been ejected from the body of the galaxy as an unevolved OB star. Note, however, that the kinematics of the SMC is quite complicated. Stanimirovic et al. (2004) found that there is a velocity gradient throughout the galaxy with the northeast quadrant (close to where this star is located) rotating up to 10 km s$^{-1}$ faster than the SMC's average heliocentric radial velocity. Thus, it is possible that this star's peculiar velocity is smaller than expected, but by no more than 10 km s$^{-1}$ (see Figure 3 in Stanimirovic et al. 2004).

We suspect that the YSG began in a binary system and was flung out into space when the primary star went supernova. Gies \& Bolton (1986) concluded that supernova explosions are not the primary mechanism behind runaway stars but most runaway stars have peculiar velocities on the order of 40-80 km s$^{-1}$ and this runaway's is much higher. As noted by Evans \& Massey (2015), only two Galactic O stars are known to have peculiar velocities $>$ 100 km s$^{-1}$, and all three runaway RSGs in the Milky Way have peculiar velocities $<$ 70 km s$^{-1}$. We believe that the high peculiar velocity of this YSG points to it originating due to a supernova explosion following the original suggestion of Zwicky (1957) and Blaauw (1956a, 1956b).

Many runaway stars are found using their bow shocks, or a presence of an infrared excess (Gvaramadze et al. 2010). We estimated that in this case the bow shock would be around 3.0$\arcsec$ away from the star. We first looked for any sign of the bow shock in MIPS SAGE-SMC data (Gordon et al.\ 2011) but sadly the resolution was too poor. We then obtained H$\alpha$ imaging of the star using the Slope 1-m telescope at Las Campanas but again we saw nothing. Finally, we looked at MCELS data (Smith et al.\ 2000) to no avail. If there is a bow shock for this runaway star, we were not able to find it.

Another interesting point to make is that these runaway stars can act as dispersing mechanism for enriched elements; for example, this YSG will eventually explode as a supernova  far away from where it was created. Since such a large fraction of OB stars are runaways, this must occur on a fairly regular basis. Thus, the process of runaway stars acts as a mechanism for dispersing elements throughout the galaxy. Although massive stars are well known to be the major source of many of the heavier elements (Maeder 1981), runaways have been largely ignored in favor of  diffusion and similar processes as a means of dispersal  (see, e.g., Pipino \& Matteucci 2009).  Recent evidence, however, suggests that our own Solar System may have been created by a runaway Wolf-Rayet star (Tatischeff \& Sereville 2010)!

As future work, we hope to look into other potential YSG runaways that we found as part of a similar project in the Large Magellanic Cloud (LMC). One star in particular, J04530398-6937285 has a radial velocity of 373 km s$^{-1}$, much higher than expected, and warrants further investigation\footnote{Another star, J04482407-7104012 (CPD-71$^\circ$285, HD 268700) has a radial velocity of 401 km s$^{-1}$, but Gaia's result results give a parallax measurement consistent with this star being a foreground halo giant at a distance of 350~pc.}.

\acknowledgements
We would like to thank Maria Drout for help collecting the discovery spectra for this star as part of the larger project to identify all YSGs in the SMC. We would also like to thank Sean Points from the MCELS team who helped us try and find evidence of a bow shock. We very much appreciate the continued support of our work from the Arizona Time Allocation Committee. Finally, we would like to acknowledge the NSF grant AST-1612874.

\clearpage

\clearpage
\begin{deluxetable}{l l l l l l l l}
\tabletypesize{\tiny}
\tablecaption{\label{tab:obs} Observations}
\tablewidth{0pt}
\startdata
\hline
Date & HJD & Telescope/Instrument & Wavelength & Resolution & Exp. time (s) & HRV (km s$^{-1}$) \\
\hline
09 Oct 2009 & 2455113.67 & CTIO 4-m/Hydra & 7300-9050\AA & 3000 & 900 & 296.4\\
16 Aug 2017 & 2457981.87 & DuPont 2.5-m/Echelle & 3500-9050\AA* & 45000 & 1800 & 298.0\\
31 Dec 2017 & 2458118.59 & Baade 6.5/MagE  & 3200\AA-1$\mu$m  & 4100 & 1800 & 305.1\\
1 Jan 2018 & 2458119.58 &  Baade 6.5/MagE  & 3200\AA-1$\mu$m  & 4100 & 600 & 305.7\\
\enddata
\tablenotetext{*}{Underexposed below 4200\AA.}
\end{deluxetable}

\begin{deluxetable}{l l l l l r}
\tablecaption{\label{tab:stds} Radial Velocities of Classification Standards}
\tablewidth{0pt}
\startdata
\hline
&                     Type  &   MagE* &   Published* &  Ref  & (MagE - Published)* \\
\hline
Sk 55        &  F3 Iab  & 168.4    &    159.9     &     1   &     8.5   \\
HD271182 & F8 0     &    309.9   &   323.9     &    1    &  -14.0\\
HD269953 & G0 0    &    259.0   &    240.4    &     1   &    18.6\\
HD269723 & G4 0   &    323.4   &    332.0    &     2   &    -8.6\\
HDE268757 &G8 0   &     284.8   &   271.6    &      1   &   13.2
\enddata
\tablenotetext{*}{units of km s$^{-1}$}
\tablenotetext{1}{Kordopatis et al. 2013}
\tablenotetext{2}{Fehrenbach \& Duflot 1982}
\end{deluxetable}

\begin{deluxetable}{l r r r c r c c r r c c c c l}
\tabletypesize{\tiny}
\rotate
\tablecaption{\label{tab:allobs} Properties of Observed Targets}
\tablewidth{0pt}
\tablehead{
\colhead{2MASS}
&\colhead{$\alpha$(J2000)}
&\colhead{$\delta$(J2000)}
&\colhead{$K$}
&\colhead{$J-K$}
&\colhead{$V$}
&\colhead{$B-V$}
&\colhead{Ref.\tablenotemark{b}}
&\colhead{Vel}
&\colhead{$r$\tablenotemark{c}}
&\colhead{Category\tablenotemark{d}}
&\colhead{Hydra}
&\colhead{Literature}
&\colhead{Ref.\tablenotemark{e}}
&\colhead{Comments\tablenotemark{f}} \\
& & & & & & & & 
\colhead{(km s$^{-1}$)}
& & &
\colhead{Class}
&\colhead{Class}
}
\startdata
J00342570-7322334               &       0       34      25.69   &       -73     22      33.5    &       11.6    &       0.25    &       12.79   &       \nodata &       3       &       -9.98   &       100.2   &       3       &       \nodata &       \nodata &       \nodata &       \nodata \\

J00342883-7328192               &       0       34      28.82   &       -73     28      19.3    &       10.72   &       0.59    &       13.03   &       0.97    &       4       &       91.12   &       109.7   &       2       &       \nodata &       G6IV    &    12  &       \nodata \\

J00344094-7319366               &       0       34      40.94   &       -73     19      36.7    &       9.18    &       0.69    &       11.73   &       \nodata &       3       &       38.02   &       106.5   &       3       &       \nodata &       \nodata &       \nodata &       \nodata \\

J00345529-7213044               &       0       34      55.30   &       -72     13      04.4    &       11.58   &       0.31    &       12.93   &       \nodata &       3       &       -10.29  &       56      &       3       &       \nodata &       \nodata &       \nodata &       \nodata \\

J00345992-7339156               &       0       34      59.91   &       -73     39      15.8    &       10.66   &       0.33    &       12.13   &       \nodata &       3       &       9.92    &       77.3    &       3       &       \nodata &       F8      &    6  &       Flo 72  \\

J00350326-7332586               &       0       35      03.25   &       -73     32      58.8    &       10.77   &       0.8     &       13.8    &       1.54    &       4       &       134.82  &       75.4    &       1       &       \nodata &       G6Ib/II &    12  &       SkKM 2  \\
\enddata
\tablenotetext{a}{These stars were imaged twice and their results were averaged. The average difference in Vel$_{\rm obs}$ was 2.2 km s$^{-1}$ and the average difference in the $r$ parameter was 13.7.}
\tablenotetext{b}{References for $B-V$ colors and $V$ magnitudes. 1 = Massey 2002; 2 = Mermilliod 1997; 3 = ASAS-3; 4 = Zaritsky 2002; 5 = (Zaritsky B) - (ASAS-3 V)}
\tablenotetext{c}{Tondry \& Davis 1979 $r$ parameter.}
\tablenotetext{d}{Category: 1 = SMC supergiant; 2 = possible SMC supergiant; 3 = foreground dwarf}
\tablenotetext{e}{References for spectral classifications. 
1 = Cannon \& Pickering 1918; 
2 = Massey \& Olsen 2003; 
3 = Evans et al.\ 2004; 
4 = Sanduleak 1969; 
5 = L\"{u} 1971; 
6 = Florsch 1972; 
7 = Lennon 1997;
8 = Azzopardi et al.\ 1975; 
9 = Houk \& Cowley 1975; 
10 = Ardeberg \& Maurice 1977; 
11 = Azzopardi \& Vigneau 1979; 
12 = Gonz\`{a}lez-Fern\`{a}dez et al.\ 2015;
13 = Humphreys 1983; 
14 = Humphreys et al.\ 1991; 
15 = Wallerstein 1984; 
16 = Bouchet et al.\ 1985; 
17 = Carney et al.\ 1985; 
18 = Sanduleak 1989}
\tablenotetext{f}{Comments include: alternate names for the star, comments on the blue spectra and comments based on the literature search.}
\end{deluxetable}

\begin{deluxetable}{l c c c c c c c c c}
 \tabletypesize{\scriptsize}
 \tablecaption{\label{tab:temps} Effective Temperatures Derived from ATLAS9 Models}
 \tablewidth{0pt}
 \tablehead{ 
 &
 &\multicolumn{2}{c}{$E(B-V)=0.04$} 
 &
 &\multicolumn{2}{c}{$E(B-V)=0.09$} 
 & 
 &\multicolumn{2}{c}{$E(B-V)=0.14$} \\ \cline{3-4} \cline {6-7} \cline {9-10} \\
 \colhead{Color} 
 & \colhead{Observed}
 & \colhead{Intrinsic}
 &\colhead{$T_{\rm eff}$ (K)}
 &
 & \colhead{Intrinsic}
 &\colhead{$T_{\rm eff}$ (K)}
 &
 & \colhead{Intrinsic}
 &\colhead{$T_{\rm eff}$ (K)} \\
 }
 \startdata
 U-B & 0.78 & 0.75 & 4906 & & 0.72 & 4943 & & 0.68 & 4993 \\
 B-V  & 1.15 & 1.11 & 4768 & & 1.06 & 4857 & & 1.01 & 4945 \\
 V-K  & 2.94 & 2.82 & 4392 & & 2.67 & 4499 & & 2.53 & 4508 \\
 J-K  & 0.78 & 0.76 & 4394 & & 0.73 & 4461 & & 0.71 & 4508 \\
 Avg.  & \nodata & \nodata & 4615 && \nodata & 4690 && \nodata & 4766\\
 \enddata
 \end{deluxetable}
 
\begin{deluxetable}{l l}
\tablecaption{\label{tab:physprop} Physical Properties}
\tablewidth{0pt}
\tablehead{
\colhead{Parameter}
&\colhead{Value}
}
\startdata
\hline
Mass & 9 M$_{\odot}$ \\
Age & 30 Myr \\
T$_{\rm eff}$ & 4700 K\\
$\log$ g [cgs] & 0.8\\
M$_{\rm V}$ & -5.4\\
$\log \frac{{\rm L}}{{\rm L}_{\odot}}$ & 4.2\\
Radius & 190 R$_{\odot}$ \\
\enddata
\end{deluxetable}


\begin{references}
\reference {} Ardeberg, A. \& Maurice, E. 1977, A\&AS, 30, 261
\reference {} Azzopardi, M. \& Vigneau, J. 1979, A\&AS, 35, 353
\reference {} Azzopardi, M., Vigneau, J. \& Macquet, M. 1975, A\&AS, 22, 285
\reference {} B\"{o}hm-Vitense, E. 1981, ARAA 19, 295
\reference {} Bouchet, P., Lequeux, J., Maurice, E., Prevot, L. \& Prevot-Burnichon, M. L. 1985, A\&A, 149, 330
\reference {} Blaauw, A. 1956a, PASP, 68, 495
\reference {} Blaauw, A. 1956b, ApJ, 123, 408
\reference {} Cannon, A. J. \& Pickering, E. C. 1918, Annals of Harvard College Observ., 91, 1
\reference {} Carney, B. W., Janes, K. A. \& Flower, P. J. 1985, AJ, 90, 1196
\reference {} Capuzzo-Dolcetta, R., \& Fragione, G. 2015, MNRAS, 454, 2677
\reference {} Castelli, F., \& Kurucz, R. L. 2004, in Modeling of Stellar Atmospheres, IAU Symp.\ 210, ed. N. E. Piskunov, W. 
\reference {} Cox, A. N. 2000, Allen's Astrophysical Quantities, 4th Edition (New York: Springer)
\reference {} Cox, N. L. J. et al.\ 2012, A\&A, 537, A35
\reference {} Ekstr\"{o}m, s., et al., 2012, A\&A, 537, 146
\reference {} Evans, C. J., Howarth, I. D., Irwin, M. J., Burnley, A. W. \& Harries, T. J. 2004, MNRAS, 353, 601
\reference {} Evans, K. A., \& Massey, P. 2015, AJ, 150, 149
\reference {} Fehrenbach, C., \& Duflot, M. 1982, A\&AS, 48, 409
\reference {} Florsch, A. 1972, Publ.\ Obs.\ Astron.\ Strausbourg, 2, 1
\reference {} Fragione, G., \& Capuzzo-Dolcetta, R. 2016, MNRAS, 458, 2596
\reference {} Georgy, C., Ekstr\"{o}m, S., Granada, A., Meynet, G., Mowlavi, N., Eggenberger, P., \& Maeder, A., 2013, A\&A, 553, 24
\reference {} Gies, D. R., \& Bolton, C. T. 1986, ApJS, 61, 419
\reference {} Gonz\`{a}lez-Fern\`{a}dez, C., Dorda, R., Negueruela, I., \& Marco, A., 2015, A\&A, 578, 3
\reference {} Gordon, K. D., et al. 2011, AJ, 142, 102
\reference {} Gvaramadze, V. V., Kroupa, P., \& Pflamm-Altenburg, J., 2010, A\&A, 519, 33
\reference {} Gvaramadze, V. V., Menten, K. M., \& Kniazev, A. Y. et al.\ 2014, MNRAS, 437, 843
\reference {} Houk, N. \& Cowley, A. P. 1975, Univ.\ of Mich.\ Cat.\ of 2D Spect.\ Types for HD Stars, C01, 0
\reference {} Humphreys, R. M. 1983, ApJ, 265, 176
\reference {} Humphreys, R. M., \& McElroy, D. B. 1984, ApJ, 284, 565
\reference {} Humphreys, R. M., Kudritzki, R. P. \& Groth, H. G. 1991, A\&A, 245, 593
\reference {} Kaler,  J. B. 2011,  Stars and Their Spectra (Cambridge: Cambridge University Press), 186
\reference {} Keenan, P. C., \& McNeil, R. C. 1989, ApJS, 71, 245
\reference {} Kordopatis, G., Gilmore, G., Steinmetz, M. et al.\ 2013, AJ, 146, 134
\reference {} Lennon, D. J. 1997, A\&A, 317, 871
\reference {} Leonard, P. J. T., \& Duncan, M. J. 1990, AJ, 99, 608
\reference {} Levesque, E. M., Massey, P., \& Olsen, K. A. G. 2006, ApJ, 645, 1102
\reference {} L\"{u}, P. K. 1971, Trans.\ of the Astron.\ Obs.\ Yale Univ., 31, 1
\reference {} Maeder, A. 1981, A\&A, 101, 385
\reference {} Massey, P. \& Olsen, K. A. G. 2003, AJ, 126, 2867
\reference {} Massey, P., Olsen, K. A. G., Hodge, P. W., Jacoby, G. H., McNeill, R. T., Smith, R. C., \& Strong, S. B. 2007, AJ, 133, 2393
\reference {} Neugent, K. F., Massey, P., Skiff, B., Drout, M. R., Meynet, G., \& Olsen, K. A. G. 2010, ApJ, 719, 1784
\reference {} Noriega-Crespo, A., van Buren, D., Cao, Y. and Dgani, R. 1997, AJ, 114, 837
\reference {} Pipino, A., \& Matteucci, F. 2009, in The Age of Stars, IAU Symp.\ 258, ed. E. E. Mamajek, D. R. Soderblom, \& R. F. G. Wyse, (Cambridge: Cambridge Univ.\ Press), 39
\reference {} Richter, O. -G., Tammann G. A. \& Huchtmeier, W. K. 1987, A\&A, 171, 33
\reference {} Sanduleak, N. 1969, AJ, 74, 877
\reference {} Sanduleak, N. 1989, AJ, 98, 925
\reference {} Schlegel, D. J., Finkbeiner, D. P., \& Davis, M. 1998, ApJ, 500, 525
\reference {} Smith, C., Leiton, R., \& Pizarro, S. 2000, ASPC, 221, 83
\reference {} Stanimirovi\'{c}, S., Staveley-Smith, L. \& Jones, P. A. 2004, ApJ, 604, 176
\reference {} Tatischeff, V., Duprat, J., \& de Sereville, N. 2010, ApJ, 714L, 26
\reference {} Tonry, J. \& Davis, M. 1979, AJ, 84, 1511
\reference {} van den Bergh, S. 2000, in The Galaxies of the Local Group (Cambridge: Cambridge University Press)
\reference {} Wallerstein, G. 1984, AJ, 89, 1705
\reference {} Zaritsky, D., Harris, J., Thompson, I. B., Grebel, E. K. \& Massey, P. 2002, AJ, 123, 855
\reference {} Zwicky, F. 1957,  Morphological Astronomy, (Berlin: Springer), 258

\end{references}
\end{document}